\documentclass[aps,twocolumn,pra,tightenlines,floatfix,showpacs,superscriptaddress]{revtex4-1}
\usepackage{graphicx}
\usepackage{amsmath}
\usepackage{amssymb}
\usepackage{times}
\usepackage[english]{babel}

\begin{document}

\title{Mean-field description of pairing effects, BKT physics, and superfluidity in 2D Bose gases}
\author{Chih-Chun Chien}
\affiliation{Theoretical Division, Los Alamos National Laboratory, Los Alamos, NM 87545, USA}
%\email{chihchun@lanl.gov}
\author{Jianhuang She}
\affiliation{Theoretical Division, Los Alamos National Laboratory,  Los Alamos, NM 87545, USA}
%\email{she@lanl.gov}
\author{Fred Cooper}
\affiliation{Theoretical Division, Los Alamos National Laboratory,  Los Alamos, NM 87545, USA}
\affiliation{Santa Fe Institute, Santa Fe, New Mexico 87501, USA}
%\email{cooper@santafe.edu}

\date{\today}

\begin{abstract}
We derive a  mean-field  description for two-dimensional (2D) interacting Bose gases at arbitrary temperatures. We find that  genuine Bose-Einstein condensation with long-range coherence only survives at zero temperature. At finite temperatures, many-body pairing effects included in our mean-field theory introduce a finite amplitude for the pairing density, which results in a finite superfluid density. We incorporate Berenzinskii-Kosterlitz-Thouless (BKT) physics into our model by considering the phase fluctuations of our pairing field. This then leads to the result that the superfluid phase is only stable below the BKT temperature due to these  phase fluctuations. In the weakly interacting regime  at low temperature we compare our theory to previous results from perturbative calculations, renormalization group calculations as well as Monte Carlo simulations. We present a finite-temperature phase diagram of 2D Bose gases. One signature of the finite amplitude of the pairing density field is a two-peak structure in the single-particle spectral function, resembling that of the pseudogap phase in 2D attractive Fermi gases. 
\end{abstract}


\maketitle

\section{Introduction}
Recent experiments on two-dimensional (2D) ultra-cold atoms have explored many interesting phenomena including the Berenzinskii-Kosterlitz-Thouless (BKT) physics \cite{HadzibabicBKT}, superfluidity \cite{Clade2D}, scale invariance \cite{Hung2D}, radio-frequency (RF) spectroscopy \cite{KohlRF}, thermodynamics \cite{Yefsah2D}, pseudogap physics above the BKT transition temperature \cite{KohlPG}, and others. These experiments provide opportunities for studying more complicated 2D or layered systems related to high-temperature superconductors \cite{PFfermion2D,Benfatto07} and interface superconductivity \cite{Caviglia08}. Theoretical studies on dilute 2D Bose gases have been reviewed in Ref.~\cite{PosazhennikovaRMP} and those available theories are confined to weakly interacting regimes or temperatures close to zero or near the critical regime. As a consequence, finite-temperature phase diagrams shown in Ref.~\cite{PosazhennikovaRMP} are schematic instead of coming from a consistent theoretical description. A mean-field theory that works in the regime of intermediate interaction strength at arbitrary temperature thus would be highly desired for a systematic analysis of 2D interacting Bose gases. 
To better understand the physics, one needs a coherent description of superfluidity, the BKT transition, pairing effects, and single-particle excitation energy. The goal of this paper is to present a plausible mean-field theory with experimental consequences for a 2D interacting single-species Bose gas.

For an attractive 2D two-component Fermi gas, there have been theories based on the phase fluctuations of the BCS theory and its extension to Bose-Einstein condensation (BEC) of dimers \cite{PFfermion2D,Melo2D}. When the temperature $T$ is below a pairing onset temperature, pairs with disordered phases emerge. When $T$ falls below the BKT transition temperature $T_{BKT}$, a superfluid phase becomes stable but a genuine long-range ordered phase only survives at $T=0$. For bosons we may explore similar physics. Several questions follow: How can BKT physics be incorporated into a theory of 2D interacting bosons? Does any interesting phase exist above $T_{BKT}$? Can 2D bosons have an energy gap in the single-particle excitation?  These issues will be addressed in a consistent theoretical framework.

Ref.~\cite{PosazhennikovaRMP} poses a series of questions that needs to be explored in theoretical work on 2D interacting Bose gases. The last one is "Could one justify a large-$N$ approach which improves on existing methods by incorporating the $t$-matrix approximation?". Inspired by this question,  here we base our theory on the leading-order-auxiliary-field (LOAF) theory of interacting bosons \cite{LOAFPRL,LOAFlong}, which is a generalization of  the conventional large-$N$ expansion \cite{MoshelargeN,Zeebook,ourlargeN2BEC}. It also reduces to the large-N expansion in the normal phase. The LOAF theory is a mean-field theory for interacting Bose gases. Its advantages in describing a 3D Bose gas beyond perturbative regimes at arbitrary temperature are summarized below. In the following we will construct a mean-field theory that applies to 2D Bose gases and explore its thermodynamics and possible experimental implications.

For 3D interacting bosons the LOAF theory meets three important criteria by treating the pairing (anomalous)  density field and the (normal)  density field on equal footing: (i) a gapless dispersion in the BEC phase, (ii) a conserving theory, and (iii) predicts a second-order BEC transition. Widely used theories such as the Hartree-Fock theory or the Popov theory fail at least one criterion \cite{LOAFlong}. Moreover, the LOAF theory exhibits a shift in the critical temperature $T_c$ consistent with the results of Ref.~\cite{Baym2000}. We emphasize that the LOAF theory naturally recovers the Bogoliubov theory of weakly interacting bosons \cite{LOAFPRL} and note that  two Green's functions corresponding to our two density fields are indeed present in the Bogoliubov theory \cite{Fetterbook}.  An important feature of the LOAF approximation is that the superfluid density is closely related to the pairing density \cite{LOAF_SF} and the two quantities obey a Josephson relation \cite{LOAF_Josephson}. This will be crucial in integrating the BKT physics into the LOAF theory of 2D Bose gases.

Before presenting our theory, we point out several challenges in developing a consistent mean-field theory for 2D interacting Bose gases. For 2D systems with short-ranged interactions, Mermin-Wagner theorem \cite{MWtheorem} rules out the possibility of long-range order. Therefore at finite $T$ a genuine  condensate is not possible in the thermodynamic limit. The lack of a genuine condensate implies that a gapless Goldstone mode may not exist so that  the dispersion of excitations should be gapped rather than having the gapless Bogoliubov dispersion relation found in three dimensions in the condensate phase. This challenge may be circumvented by confining the discussion on finite-size patches of the system or introducing phase fluctuations to disorder the condensate \cite{PosazhennikovaRMP}. Here we consider 2D Bose gases in the thermodynamic limit so we follow the second approach. The two-body scattering problem in 2D is also very different from that in 3D (see Refs.~\cite{2Dscattering,PosazhennikovaRMP}). The scattering amplitude in 3D is finite as both the energy- and momentum- transfers approach zero. This allows one to develop an effective theory with a coupling constant set equal to the scattering amplitude at zero energy- and momentum- transfer. In contrast, the 2D scattering amplitude vanishes in the limit of zero energy- and momentum- transfer. Following the standard renormalization methods \cite{Zeebook}, one has to formulate a running coupling constant which should be defined at a finite energy (or momentum) scale. A renormalization scheme for the running coupling constant will be presented and the running coupling constant indeed exhibits behavior consistent with the $t$-matrix analysis of the scattering amplitude. 

This paper is organized as follows. Section \ref{sec:2DLOAF} shows the derivation of the 2D LOAF theory. The renormalization of the finite temperature effective potential and the incorporation of phase fluctuations and the BKT physics are addressed in detail. In Section \ref{sec:results} we present  the calculated  phase diagram and the  experimental implications of the 2D LOAF theory. Comparisons with perturbative calculations are also presented. Section \ref{sec:conclusion} states the conclusions of our work. 

\section{LOAF theory of 2D interacting Bose gases}\label{sec:2DLOAF}
The action of a homogeneous 2D Bose gas is given by $S=\int dx\mathcal{L}$, where $dx\equiv dt d^{2}x$ and the Lagrangian density is
\begin{equation}\label{eq:S}
\mathcal{L}=\frac{1}{2}[\phi^{*}(x)h\phi(x)+\phi(x)h^{*}\phi^{*}(x)]-\frac{\lambda}{2}|\phi(x)|^{4}.
\end{equation}
Here $h=i\hbar\partial_{t}+\hbar^{2}\nabla^{2}/2m+\mu$ and $\mu$ is the chemical potential. We set $\hbar\equiv 1$. $\lambda$ is the bare 2D repulsive coupling constant. After discussing its renormalization, we will show its connection to the 2D $s$-wave scattering length. This action is equivalent to the  Hamiltonian
\begin{equation}
H=\frac{\hbar^2}{2m}|\nabla \phi|^{2}+\frac{\lambda}{2}|\phi|^4.
\end{equation}
Although the Hamiltonian approach to BECs is more common in the literature, having a path-integral formulation of the problem allows for 
certain expansion methods that would be hard to implement in the canonical formalism.  The expansion used here is one such expansion which allows one to interpret a mean-field theory as the first term in a complete resummation of the original theory by utilizing a
Hubbard-Stratonovich transformation \cite{Hubbard,*Stratonovich} and then reversing the order of integrations. Introducing auxiliary fields allows one to 
do the path integration over the original fields exactly while holding the auxiliary fields constant. Then one does the remaining
path integrations by the method of steepest descent and obtains a loop expansion in terms of the auxiliary-field propagators.

In utilizing the Hubbard-Stratonovich transformation we impose the following requirements: (i) to treat the normal and anamolous densities on an equal footing and (ii) to reduce in the weak-interaction limit to Bogoliubov's theory of interacting bosons (see \cite{AndersenReview,Pethickbook} for reviews). To insure this latter feature  we introduce the normal and pairing density composite fields $\chi_0$ and $A$ representing $\sqrt{2}\lambda\phi^{*}(x)\phi(x)$ and $\lambda\phi(x)\phi(x)$ with the corresponding fluctuations, the Lagrangian density in the LOAF theory becomes \cite{LOAFPRL,LOAFlong}
\begin{eqnarray}\label{eq:L}
\mathcal{L}&=&\mathcal{L}_{0}+[A(x)[\phi^{*}(x)]^{2}+A^{*}(x)[\phi(x)]^{2}]- \nonumber \\
& &\sqrt{2}\chi_{0}(x)|\phi(x)|^{2}+\frac{1}{2\lambda}[\chi^{2}_{0}(x)-|A(x)|^{2}].
\end{eqnarray}
Here $\mathcal{L}_{0}$ denotes the kinetic energy part of Eq.~\eqref{eq:S}. Note that the pairing density field emerges from the pairing channel in Eq.~\eqref{eq:S}. These auxiliary fields introduce composite-field propagators and the counting of their loops facilitates a resummation scheme similar to the large-$N$ expansion \cite{LOAFlong,MoshelargeN,ourlargeN2BEC,Zeebook}.

The leading-order auxiliary-field (LOAF) theory of 3D Bose gases has been discussed in Ref.~\cite{LOAFPRL,LOAFlong} and here we briefly review its derivation for a 2D Bose gas. We begin with the Lagrangian density after introducing the composite fields $\chi_0$ and $A$ as shown in Eq.~\eqref{eq:L}. To simplify our expressions, we define $\Phi=(\phi,\phi^*,\chi_0,A,A^*)^{T}$ and its corresponding source term $J=(j,j^*,s,\tilde{S},\tilde{S}^*)^{T}$. The generating functional $W[J]$ can be obtained from the partition function by
\begin{eqnarray} \label{zj}
Z[J]&=&e^{iW[J]/\hbar}=\mathcal{N}\int\mathcal{D}\Phi e^{(i/\hbar)[S[\Phi]+\int J^{\dagger}\Phi]}, \nonumber \\
S[\Phi]&=&\int dtd^{2}x\mathcal{L}.
\end{eqnarray}

We note here that if we first perform the Gaussian path integration over the auxiliary fields $\chi_0, A, A^*$, we recover the
original generating functional based on the Lagrangian density \eqref{eq:S}.  Instead we will reverse the orders of integrations and
perform the (now) Gaussian integration over the fields $\phi, \phi^*$.  This is the gist of the Hubbard-Stratanovich transformation. The generating functional of one-particle irreducible (1-PI) graphs $\Gamma[\Phi]$ is the Legendre transformation of $W[J]$:
\begin{equation}
\Gamma[\Phi]=\int d^{d}x J^{\dagger}(x)\Phi(x)-W[J].
\end{equation}
The action $S[\Phi]$ in Eq.(\ref{zj}) written in terms of the auxiliary fields $A$ and $\chi_0$ is quadratic in the fields $\phi,\phi^\star$, so that the Gaussian integral in $\phi, \phi^*$ can be done exactly, leaving an effective action $S_{eff}[\chi_0, A, A^\star]$ which now depends on $J, \chi_0, A, A^*$. In order to perform the remaining integrals we introduce a small parameter $\epsilon$ into the problem by replacing $S_{eff} \rightarrow S_{eff}/\epsilon$.  This allows us to perform the remaining integrals via Laplace's method or the stationary-phase approximation. As shown in Refs.~\cite{LOAFlong,CooperAnnPhys}, the parameter $\epsilon$ counts the number of composite-field ($\chi$ and $A$) propagator loops. The LOAF approximation is the leading order in $\epsilon$ and consists of just keeping the contribution at the stationary phase point. The details of this expansion are given in Ref.~\cite{LOAFlong}.

The Legendre transform of the stationary phase approximation for $Z[J]$ is then found to be
\begin{eqnarray}
\Gamma[\Phi]&=&\frac{1}{2}\int d^{d}x d^{d}x^{\prime}\phi^*_{a}G^{-1}_{ab}(x,x^{\prime})\phi_{b}(x^{\prime}) \nonumber \\
& &-\int d^{d}x\left(\frac{\chi_0^2-|A|^2}{2\lambda}-\frac{\hbar}{2i}\mbox{Tr}\{\ln[G^{-1}(x,x)]\} \right). \nonumber
\end{eqnarray}
Here $G^{-1}(x,x^{\prime})=\delta(x,x^{\prime})\tilde{G}^{-1}$ and $\tilde{G}^{-1}$ is given by
\begin{equation}
\left(\begin{array}{cc} 
-i\hbar\partial_t-\hbar^2\nabla^2/(2m)+\chi & -A \\
-A^* & i\hbar\partial_t-\hbar^2\nabla^2/(2m)+\chi
\end{array} \right). \nonumber
\end{equation}
Here $\chi=\sqrt{2}\chi_0-\mu$. The effective potential can be evaluated by a Wick rotation to the imaginary time $\tau\rightarrow it$ and using the standard Matsubara frequency summation. For homogeneous static fields we define the effective potential $V_{eff}\equiv \Gamma[\Phi]/\mathcal{V}\beta$, where $\mathcal{V}$ is the volume of the system and $\beta=1/(k_{B}T)$. 
The inverse Green's function becomes
\begin{equation}
G^{-1}=\left(\begin{array}{cc} 
-i\omega_n+\epsilon_k+\chi & -A \\
-A^* & i\omega_n+\epsilon_k+\chi
\end{array} \right). \nonumber
\end{equation}
Here $\omega_n$ is a bosonic Matsubara frequency and $\epsilon_k=\hbar^{2}k^{2}/(2m)$. The last term in $V_{eff}$ is $\frac{1}{2\beta}\mbox{Tr}\{\ln G^{-1}\}=\frac{1}{2}\sum_{k}\frac{1}{\beta}\sum_{n}\ln(\omega_n^{2}+\omega_{k}^{2})=\sum_{k}[(\omega_k/2)+(1/\beta)\ln(1-e^{-\beta\omega_k})]$. Here $\omega_{k}=\sqrt{(\epsilon_k+\chi)^2-|A|^2}$. Then one obtains the expression for the (unrenormalized) effective potential 
\begin{eqnarray}\label{eq:V_unrenorm}
V_{eff}&=&\chi|\phi|^2-\frac{A^*\phi^2}{2}-\frac{A(\phi^*)^2}{2}-\frac{(\chi+\mu)^2}{4\lambda}+\frac{|A|^2}{2\lambda}+ \nonumber \\
& &\sum_{k}\left[\frac{\omega_k}{2}+\frac{1}{\beta}\ln (1-e^{-\beta\omega_k}) \right].
\end{eqnarray}

At the minimum of the potential  $\delta V_{eff}/\delta \phi^*=0$ which leads to the condition $\chi\phi-A\phi^*=0$.  Thus if the minimum is not at $\phi=0$, we can use the $U(1)$ symmetry to make $\phi$ and $A$  real and then we obtain in that case $\chi=A$. 
Indeed, at $T=0$ the minimum is not at zero and  the single-particle Bose-Einstein condensation (BEC) corresponds to a finite $\phi=\sqrt{\rho_0}$. Therefore in the presence of the single-particle BEC, $\chi=A$ and the dispersion is gapless $\omega_k=\sqrt{\epsilon_k(\epsilon_k+2\chi)}$, where $\chi$ plays the role of $\lambda\rho_0$ in the Bogoliubov theory of weakly interacting bosons.

\subsection{Renormalization of LOAF theory}
So far the theory has not been renormalized and it is ultraviolet divergent. In Popov's approach to 2D Bose gases \cite{Popovbook,PosazhennikovaRMP} a high-energy cutoff has been introduced and this cutoff remains in the equations of state. Here we systematically renormalize the coupling constant, chemical potential, and vacuum energy so that there is no ultra-violet divergence in our theory. By examining $V_{eff}$ shown in Eq.~\eqref{eq:V_unrenorm} one can see that the ultraviolet divergence comes from the integral of $\omega_k$. The ultraviolet divergent part of $\omega_k$ at large $k$ is $\epsilon_k+\chi-|A|^2/(2\epsilon_k)$ in both 3D and 2D.  In 3 D one can relate these terms to usual renormalizations evaluated 
at $\Omega = q=0$, where $\hbar\Omega$ and $q$ are the energy transfer and momentum transfer of two-body scattering.  In two dimensions, however, one needs to define the renormalized coupling constant at a finite value of $\Omega$ and/or $q$ since the scattering amplitude vanishes at zero $\Omega$ and $q$  \cite{2Dscattering,PosazhennikovaRMP}.

In our resummed theory the renormalized running coupling constant is defined as the inverse of the second derivative of the effective action with respect to the field $A$.
This leads to an equation
\begin{eqnarray}
\frac{1}{\lambda_R(q,\Omega, T)}  &&= 2 D^{-1}_{AA^*}  (q,\Omega, T). 
\end{eqnarray}
The inverse $AA^*$ propagator can be obtained by $\delta^2 \Gamma[\Phi]/\delta A \delta A^*$. Explicitly,
\begin{eqnarray}
D^{-1}_{AA^*}  (x,x^{\prime}, T)&&= \frac{\delta(x-x^{\prime})}{2\lambda} + \frac{1}{2}\frac {\delta^2\mbox{Tr} \ln[ G^{-1}] } {\delta A(x) \delta A^{*}(x^\prime)}.
\end{eqnarray}
After a Fourier transform this becomes
\begin{eqnarray}
D^{-1}_{AA^*}  (q,\Omega, T)&&= \frac{1}{2\lambda} - \frac{1}{2}\Pi (q,\Omega, T).
\end{eqnarray}
Here $\Pi (q,\Omega, T)$ is the bubble diagram discussed in Ref.~\cite{LOAF_SF}. The divergent part of $\Pi (q,\Omega=0, T=0)$ in 2D is $(1/2\pi)\int dk k [\sqrt{\epsilon_k^2+(\epsilon_q/2)^2}]^{-1}$.
 We  define the physical renormalized coupling constant  in 2D as the running coupling constant at $T=0, \Omega = 0, q=q_0$
(of course we can choose any other scale to define the renormalized coupling constant).   For a comparison with the relativistic case, 
one can refer to Ref.~\cite{CJP} and references therein. 

We define an intermediate  renormalized coupling constant by just keeping  the ultraviolet divergent part of the bubble diagram.
\begin{equation}\label{eq:lambda2D}
\frac{1}{\lambda_R(q_0)}=\frac{1}{\lambda}-\int\frac{dkk}{2\pi}\frac{1}{\sqrt{\epsilon_k^2+(\epsilon_{q_0}/2)^2}}.
\end{equation}
In terms of this intermediate renormalized  coupling constant (which we will choose to be our definition of the renormalized coupling constant $\lambda_R$ )   one finds  that the running coupling constant at any scale is given by 
\begin{eqnarray}
\frac{1}{\lambda_R(q,\Omega, T)} &=& \frac{1}{\lambda_R(q_0) }  -\left( \Pi(q,\Omega, T)  - \right. \nonumber \\
& &\left. \int\frac{dkk}{2\pi}\frac{1}{\sqrt{\epsilon_k^2+(\epsilon_{q_0}/2)^2}}\right).
\end{eqnarray}
In three dimensions one is allowed to set $q_0 = 0$, since the two-body scattering is finite at zero momentum transfer.

The connection between the 2D renormalized running coupling constant and the $t$-matrix calculations of the 2D scattering amplitude can be seen clearly at $T=0$. One can show that $\Pi(q,\Omega=0, T=0)  - \Pi(q_0,\Omega=0, T=0) = (1/2\pi)\ln(q/q_0)$ and as a consequence
\begin{eqnarray}
\lambda_{R}(q,\Omega=0, T=0)=\frac{\lambda_{R}(q_0)}{1-\frac{1}{2\pi}\ln\left(\frac{q}{q_0}\right)}.
\end{eqnarray}
Therefore $\lambda_R(q,\Omega=0,T=0)\rightarrow 0$ as $q\rightarrow 0$, which is consistent with the finding that the 2D scattering amplitude vanishes at zero energy- and momentum- transfer \cite{2Dscattering,PosazhennikovaRMP} in the $t$-matrix calculations. This analysis justifies our choice of a finite momentum transfer in defining the running coupling constant. The 2D scattering amplitude at a small momentum transfer is given by $f^{2D}\sim 4\pi/[m|\ln(\rho a^2)|]$, where $a$ is the 2D $s$-wave scattering length \cite{2Dscattering,PosazhennikovaRMP}. After the coupling-constant renormalization, $\lambda_{R}$ may be set to be proportional to $f^{2D}$ and we formally write 
\begin{equation}\label{eq:lambdaR}
\lambda_R=\frac{4\pi \hbar^2}{m|\ln(\rho a^2)|}.
\end{equation}
Here $a$ is the scattering length measured at the momentum transfer $q_0$ and $T=0$. We focus on the regime $\rho a^2 <1$, which is consistent with the requirement of a dilute gas. This less restrictive condition, when compared to the condition $|\ln|\ln(\rho a^2)||\ll 1$ for those perturbative methods reviewed in Ref.\cite{PosazhennikovaRMP}, allows us to explore intermediate-coupling regimes where $|\ln(\rho a^2)|^{-1}\sim O(1)$.

In contrast to the renormalization of the coupling constant, the chemical-potential and vacuum-energy renormalizations are found to have the same form   in  two and three dimensions. The chemical-potential renormalization is 
\begin{equation}
\frac{\mu_R}{\lambda_R}=\frac{\mu}{\lambda}-\sum_{k}1,
\end{equation}
The vacuum-energy renormalization is given by 
\begin{equation}
V_R-\frac{\mu_{R}^2}{4\lambda_R}=V_0-\frac{\mu^2}{4\lambda}-\sum_{k}\frac{\epsilon_k}{2},
\end{equation}
Here $V_0$ and $V_R$ are the bare and renormalized vacuum energy. 

The LOAF  theory written in terms of the renormalized parameters is thus given by
\begin{eqnarray}
V_{eff}&=&\chi|\phi|^2-\frac{A^*\phi^2}{2}-\frac{A(\phi^*)^2}{2}-\frac{(\chi+\mu)^2}{4\lambda}+\frac{|A|^2}{2\lambda}+ \nonumber \\
& &\frac{1}{2\pi}\int_{0}^{\infty}dk k\left[\frac{\omega_k-\epsilon_k-\chi+\frac{|A|^2}{2\sqrt{\epsilon_k+(\epsilon_{q_0}/2)^2}}}{2}+ \right. \nonumber \\
& &\left. \frac{1}{\beta}\ln (1-e^{-\beta\omega_k}) \right]. \nonumber \\
& &
\end{eqnarray}
Here we drop the index $R$ and the vacuum energy and use continuity to extend our renormalization procedure to the region where $\chi \neq A$.  The equations of state are given by $\delta V_{eff}/\delta\chi=0$, $\delta V_{eff}/\delta A^*=0$, and $-\delta V_{eff}/\delta \mu=\rho$, which are explicitly shown in the main text.

The LOAF theory then only keeps the leading order terms 
 and one obtains the equations of state (EOS) by minimization of the renormalized $V_{eff}$:
\begin{eqnarray}\label{eq:EOS}
\frac{A}{\lambda}&=&\rho_{0}+A\int\frac{d^{2}k}{(2\pi)^{2}}\left[\frac{1+2n(\omega_{k})}{2\omega_{k}}-\frac{1}{2\sqrt{\epsilon_k^2+(\epsilon_{q_0}/2)^2}}\right]; \nonumber \\
\rho&=&\rho_{0}+\int\frac{d^{2}k}{(2\pi)^{2}}\left[\frac{\epsilon_{k}+\chi}{2\omega_{k}}[1+2n(\omega_{k})]-\frac{1}{2} \right].
\end{eqnarray}
Here $\chi=\sqrt{2}\chi_0-\mu$, $\omega_k=\sqrt{(\epsilon_k+\chi)^{2}-|A|^{2}}$, $n(x)=1/(e^{\beta x}-1)$ is the Bose distribution function, and $\rho_0=\phi_0^{2}$ is the condensate density, where $\phi_0$ denotes the expectation value of $\phi$. 
The density is related to the chemical potential via $\rho=(\chi+\mu)/2\lambda$. In addition to the EOS, there is a BEC condition 
$\chi\phi-A\phi^{*}=0$. 
We define $k_{0}^{2}=\rho$, $E_0\equiv\hbar^{2}k_{0}^{2}/2m\equiv k_{B}T_0$ and use $k_0$, $E_0$, and $T_0$ as our units with $k_{B}\equiv 1$. The quantity $q_0$ is the scale that one defines the running coupling constant $\lambda(q,\Omega,T)$ in a 2D Bose gas. It also defines the momentum transfer at which the scattering length $a$ is measured. We emphasize again that in 2D $q_0$ cannot be chosen to be zero due to the peculiar behavior of the scattering amplitude. However, this choice of scale does not effect any physically measurable quantities as argued in standard renormalization methods \cite{Zeebook}. Here we choose $q_0/k_0=0.1$. To compare with experimental parameters, one only needs to find out the momentum transfer $q_0$ at which the relation $f^{2D}\sim4\pi/[m|\ln(\rho a^2)|]$ is defined. That $q_0$ then serves as the renormalization energy scale in the EOS.

\begin{figure}
  \includegraphics[width=2.7in,clip] {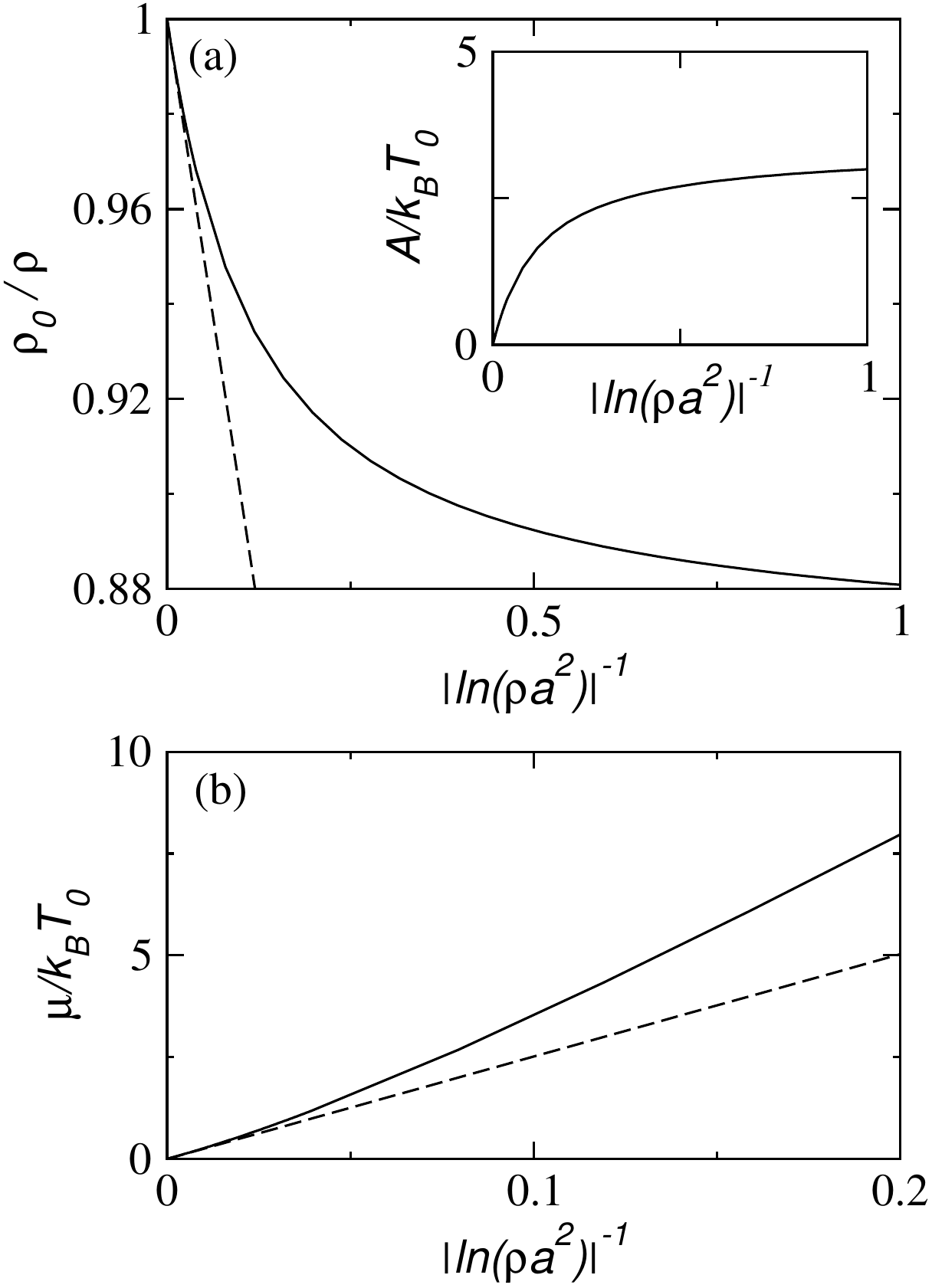}
  \caption{(a) The condensate fraction as a function of $|\ln(\rho a^2)|^{-1}$ at $T=0$ (solid line). The dashed line shows the prediction from Eq.~\eqref{eq:rho0_pert}.  The inset shows the pairing density field $A$ as a function of $|\ln(\rho a^2)|^{-1}$. (b) The chemical potential as a function of $|\ln(\rho a^2)|^{-1}$ at $T=0$ (solid line). The dashed line shows the prediction from Eq.~\eqref{eq:mu_pert}. }
\label{fig:T0phase}
\end{figure}  
At $T=0$ the LOAF theory predicts a finite $\rho_0$ and BEC. Moreover, the BEC condition $\chi\phi-A\phi^{*}=0$  requires that $\chi=|A|$ so that $\omega_k=\sqrt{\epsilon_k(\epsilon_k+2\chi)}$ is gapless. This gapless excitation is associated with the Goldstone mode in the BEC phase \cite{LOAF_SF}. The equations of state become $(A/\lambda)=\rho_0-(A/8\pi)\ln(2A/\epsilon_{q_0})$ and $\rho=\rho_0+A/8\pi$. Figure~\ref{fig:T0phase} (a) shows that $\rho_0/\rho$ decreases and $A/k_{B}T_0$ increases as the interaction $|\ln(\rho a^2)|^{-1}$ increases. The chemical potential is shown in Figure~\ref{fig:T0phase} (b). The condensate fraction and the chemical potential from perturbative calculations \cite{Schick71} are given by
\begin{eqnarray}
\mu&=&\frac{4\pi\hbar^2\rho}{m|\ln(\rho a^2)|}[1+O(|\ln(\rho a^2)|^{-1})],\label{eq:mu_pert} \\
\frac{\rho_0}{\rho}&=&1-\frac{1}{|\ln(\rho a^2)|}+O(|\ln(\rho a^2)|^{-2}). \label{eq:rho0_pert}
\end{eqnarray}
The dashed lines in Figure~\ref{fig:T0phase} show the results from perturbative calculations. One can see that for extremely small $|\ln(\rho a^2)|^{-1}$, the LOAF theory agrees well with the perturbative calculations. Thus the LOAF theory agrees quantitatively with perturbation theory  results  and should be reliable in the weakly interacting regime. Importantly, the LOAF theory could help explore interesting physics beyond the weakly interacting regime.

\subsection{Phase fluctuations}
At finite $T$, Mermin-Wagner theorem \cite{MWtheorem} rules out the possibility of long-range orders so BEC cannot survive. In other words, the $U(1)$ symmetry of the Lagrangian density \ref{eq:S} should remain unbroken at any finite $T$. The lack of a broken symmetry implies that a gapless Goldstone mode should not exist and the excitation should be gapped. This is consistent with Eq.~\eqref{eq:EOS} since the equations cannot be satisfied by a gapless dispersion. In the LOAF theory this indicates that the BEC condition cannot be met so $\rho_0$ must vanish. Eq.~\eqref{eq:EOS}, however, does not rule out the possibility of a finite $A$ and indeed we found finite values of $A$ at finite $T$. The finite expectation value of $A$ implies a diatomic condensate, which also could break the $U(1)$ symmetry of the Lagrangian density \eqref{eq:S} \cite{LOAF_SF} and violate the Mermin-Wagner theorem.

This apparent conflict was analyzed in great detail by Witten  \cite{Witten2D} in the $SU(N)$ Thirring model at large-$N$. His arguments would lead here to the conclusion  that there is a gapless mode associated with the phase
of $A$ but that this will {\it not} be a Goldstone mode. He also argues that apart from the fact that there is no symmetry breaking at finite $T$,  the fermions  (which in our case will be the boson $\phi$) will have an energy gap in the phase-disordered regime and that apart from the details of the  BKT phase transition \cite{BKT}, the large-$N$ predictions (which are similar in spirit to the LOAF theory) are quite reliable.  

In the BCS theory of attractive fermions, the method of obtaining a BKT transition in 2D by introducing phase fluctuations to disorder the would-be order parameter has been discussed in Refs.~\cite{PFfermion2D,Melo2D}. There after obtaining the BCS theory at the mean-field level, one further introduces a phase for the gap which is the analogue of the  anomalous condensate present in the LOAF theory. It has been shown that the BCS theory can be derived from the same LOAF framework \cite{LOAF_BCS} and one can obtain similar Josephson relations in both the BCS theory of attractive fermions and the LOAF theory of repulsive bosons \cite{LOAF_Josephson}. Therefore phase fluctuations may be incorporated into the LOAF theory of bosons using a method similar to that for the BCS theory of attractive fermions.
 
Following the phase-fluctuation method of the Thirring model and the BCS theory, we introduce phase fluctuations into the solution of the EOS of the LOAF theory \cite{Melo2D, Witten2D,PFfermion2D}. This procedure also introduces the BKT transition to our theory and determines where the superfluid phase is stable. The idea is to include a fluctuating phase in the pairing density field so it becomes $Ae^{i\theta}$. As discussed in Ref.~\cite{LOAF_Josephson}, the phase of the pairing field is twice of the phase of the bosonic field and both phases originate from the $U(1)$ symmetry of the Lagrangian density \eqref{eq:S}. The amplitude $A$ is determined by the EOS, Eq.~\eqref{eq:EOS}, and can be finite. Similar to Ref.~\cite{PFfermion2D} we work with the minimum where $A$ is constant. Following Refs.~\cite{Witten2D,PFfermion2D}, in the action containing $\theta$ we only keep the leading-order contribution of the phase fluctuation, which is proportional to $\int d^2x (\nabla \theta)^{2}$. This kinetic-energy term of $\theta$ corresponds to the XY model and thus the conventional BKT physics \cite{BKT} is introduced to the theory following the arguments shown in Section 5 of Ref.~\cite{PFfermion2D}.  The proportionality coefficient of the kinetic-energy term of $\theta$ is the phase stiffness, which is equal to the superfluid density $\rho_s=\delta^{2} V_{eff}/\delta v\delta v$ following a similar calculation of Ref.~\cite{LOAF_Josephson}. Here $v=\nabla\theta$.

Similar to the results presented in Ref.~\cite{PFfermion2D}, the phase fluctuations obey $\langle e^{i\theta}\rangle=0$ and as a consequence, there is no long-range coherence of the pairing density field. The correlation of the phase is $\langle e^{i\theta(\mathbf{x})}e^{i\theta(0)}\rangle \sim |\mathbf{x}|^{-mT/(2\pi\rho_{s})}$ below the BKT transition temperature $T_{BKT}$ and $\langle e^{i\theta(\mathbf{x})}e^{i\theta(0)}\rangle \sim e^{-|\mathbf{x}|/x_0}$ above $T_{BKT}$, where $x_0$ is the characteristic length for the decay of the correlation \cite{BKT}. The conventional picture is that vortex-antivortex pairs are bound below $T_{BKT}$ and a superfluid phase is possible. Above $T_{BKT}$ vortices and antivortices unbind and proliferate so the superfluid is destroyed. Thus the phase fluctuations introduce a BKT transition separating a low-$T$ superfluid phase and a non-superfluid phase.

\section{Results and Discussions}\label{sec:results}
\subsection{Finite-$T$ phase diagram}
\begin{figure}
  \includegraphics[width=2.7in,clip] {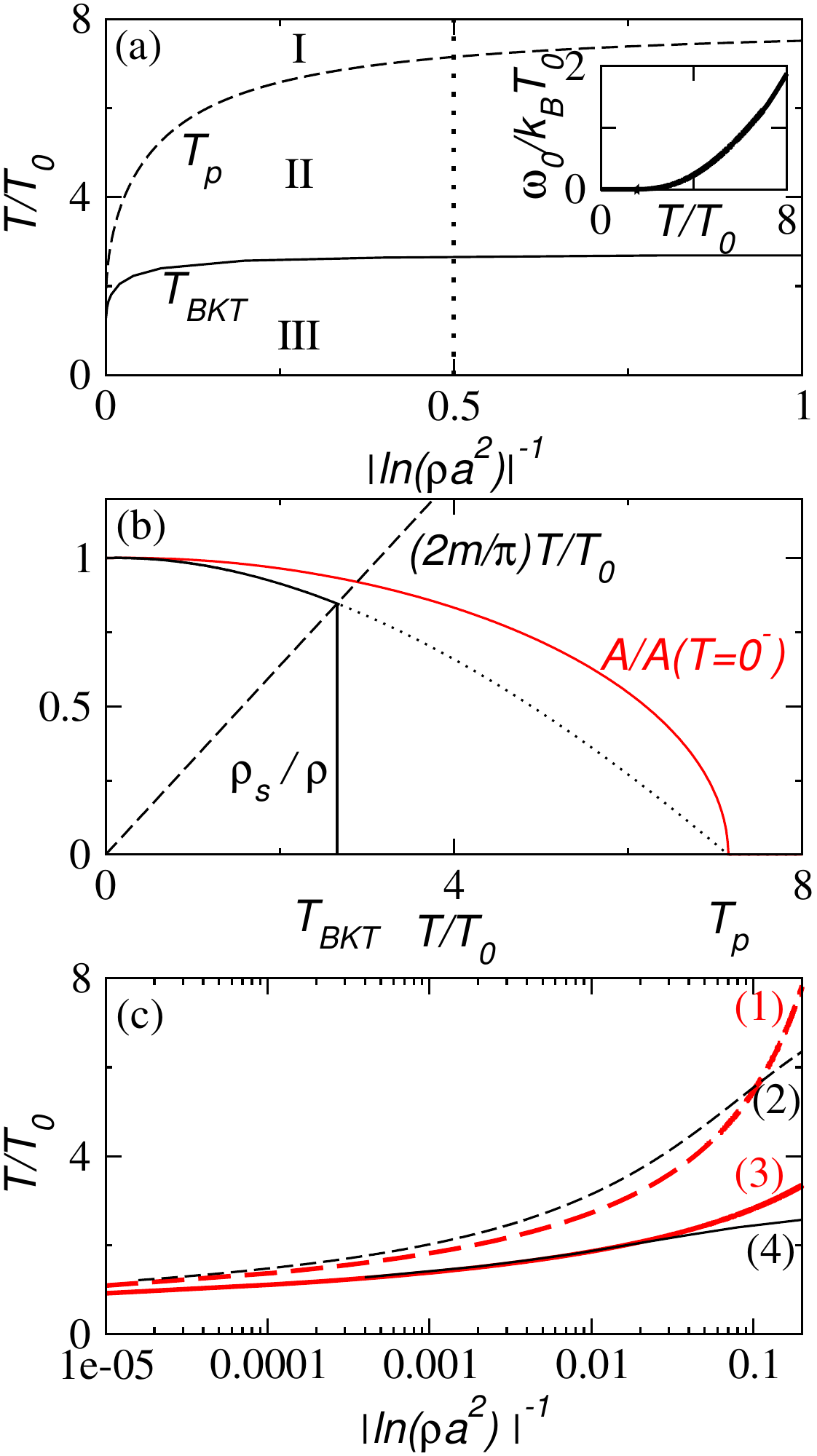}
  \caption{(a) Phase diagram of a 2D Bose gas from the LOAF theory with phase fluctuations. The solid and dashed lines show the BKT and pairing onset temperatures. Regions I, II, and III correspond to the normal, pairing, and superfluid phases. The details of physical quantities along the dotted line ($|\ln(\rho a^2)|^{-1}=0.5$) are shown in panel (b). Inset: The gap in the dispersion, $\omega_k(k=0)=\omega_0$, as a function of $T$ for $|\ln(\rho a^2)|^{-1}=0.5$. (b) The amplitude of the pairing density field $A$ and the superfluid fraction $\rho_s/\rho$ as a function of $T$ for $|\ln(\rho a^2)|^{-1}=0.5$. The dashed line is $(2m/\pi)T/T_0$ and when it intersects the curve of $\rho_s/\rho$, the BKT transition occurs and the superfluid density drops to zero. (c) Comparison of (1) $T_c^{RG}$ from Eq.~\eqref{eq:TcRG} , (2) $T_{p}$ from the LOAF theory, (3) $T_{BKT}$ from Eq.~\eqref{eq:TBKT_MC} with $C=2.152$, and (4) $T_{BKT}$ from the LOAF theory in the weakly interacting regime.}
\label{fig:2Dphase}
\end{figure} 
We now construct the finite-$T$ phase diagram. The pairing onset temperature $T_p$ is determined by the EOS \eqref{eq:EOS} when the amplitude $A$ first becomes finite. Figure~\ref{fig:2Dphase} (a) shows $T_p$ as a function of $\eta$. There is no genuine phase transition across $T_p$. Below $T_p$ bosons form composite pairing density field but the phase is random. In other words, below  $T_p$ there is a \textit{phase-disordered diatomic quasi-condensate}. We note that in the $2+1$-dimensional relativistic $O(N)$  $\phi^4$ model, there is a resonance in the $\phi-\phi$ scattering amplitude near twice the mass of the scalar meson at small coupling  \cite{CJP}.  Further studies may determine if there is finite two-body binding energy. Fig.~\ref{fig:2Dphase} (b) shows the growth of the amplitude of $A$ as $T$ decreases.

Next we investigate where superfluidity becomes stable. 
The method for determining the superfluid transition temperature comes from the analogy with the XY model once one adds the phase fluctuations \cite{PFfermion2D}. 
According to the theory of BKT transition \cite{BKT}, the superfluid is unstable above $T_{BKT}$ due to vortex-antivortex proliferation, where $T_{BKT}$ is determined by 
\begin{equation}\label{eq:TBKT}
k_{B}T_{BKT}=\frac{\pi\hbar^{2}}{2m}\rho_{s}(T_{BKT}).
\end{equation}
The superfluid density $\rho_s$ of the LOAF theory has been discussed in Ref.~\cite{LOAF_SF} and it can be obtained from the Landau two-fluid model or the current-current response function. It has been argued that the pairing density field is crucial in sustaining a finite $\rho_s$. This feature is similar to the fermionic BCS theory, where the superfluidity comes from Cooper pairs. When $\rho_0=0$ but $A>0$, one has \cite{LOAF_SF}
\begin{equation}
\rho_s=\rho-\frac{\hbar^2}{m}\int \frac{d^{2}k}{(2\pi)^{2}}\left(\frac{k^2}{2}\right)\left(-\frac{\partial n(\omega_k)}{\partial \omega_k}\right).
\end{equation}
Fig.~\ref{fig:2Dphase} (b) shows $\rho_s/\rho$ as a function of $T$ for $|\ln(\rho a^2)|=0.5$. When this curve intersects the line of $(2m/\hbar^{2}\pi)T/T_0$, the BKT condition \eqref{eq:TBKT} is met and vortex-antivortex proliferation will destroy the superfluidity above $T_{BKT}$. As a consequence, $\rho_s$ jumps to zero and the superfluid phase is only stable below $T_{BKT}$.

Following this procedure we show $T_{BKT}$ as a function of $|\ln(\rho a^2)|^{-1}$ in Fig.~\ref{fig:2Dphase} (a). There is a genuine phase transition across $T_{BKT}$ since the superfluid density is discontinuous across this boundary. We therefore identify three different phases of a 2D Bose gas at finite $T$ as shown on Fig.~\ref{fig:2Dphase} (a): Regime I above $T_p$ corresponds to a normal gas with no pairing density nor superfluidity. Regime II in between $T_{BKT}$ and $T_p$ is a non-superfluid phase with a finite amplitude of the pairing density field but no phase coherence. Regime III below $T_{BKT}$ is a superfluid phase with algebraically decaying phase correlations.  One can see that $T_p$ is about $2-3$ times of $T_{BKT}$ and is not a high-temperature scale as one may inferred from Ref.~\cite{HadzibabicBKT}. In the LOAF theory, the phase diagrams at fixed values of $|\ln(\rho a^2)|^{-1}$ are similar to the one shown in Fig.~\ref{fig:2Dphase} (b) (i. e., the amplitude of $A$ smoothly approaches zero at $T_p$ while $\rho_s$ drops to zero at $T_{BKT}$) but with different values of $T_p$ and $T_{BKT}$ which can be inferred from Fig.~\ref{fig:2Dphase}. (a).

The superfluid transition temperature in the weakly interacting regime has been theoretically studied in previous work (see \cite{PosazhennikovaRMP} for a review). In Ref.~\cite{Fisher88} a renormalization-group approach predicts that 
\begin{equation}\label{eq:TcRG}
T_{c}^{RG}\sim \frac{2\pi\rho}{m\ln|\ln(\rho a^2)|}.
\end{equation}
Monte Carlo simulations in Refs.~\cite{Prokofev01,Prokofev02} suggest the functional form
\begin{equation}\label{eq:TBKT_MC}
T_{BKT}=\frac{2\pi\rho}{m}\frac{1}{C+\ln|\ln(\rho a^2)|}.
\end{equation}
Fitting $T_{BKT}$ from the LOAF theory at weak coupling to the functional form of Eq.~\eqref{eq:TBKT_MC}  leads to the results shown in Figure~\ref{fig:2Dphase}. From the fitting we obtain $C=2.152$ as
compared with the  Monte Carlo result of Ref.~\cite{Prokofev02}, which yielded the numerical value $C= 3.409$  \footnote{In the Monte Carlo simulations of Ref.~\cite{Prokofev02},  
$C =  \ln (\xi/4\pi)$ with $\xi = 380 \pm 3$}. 
Figure~\ref{fig:2Dphase}(c) shows $T_p$ and $T_{BKT}$ from the LOAF theory along with $T_{c}^{RG}$ and $T_{BKT}$ from Eq.~\eqref{eq:TBKT_MC} with a constant $C=2.152$ in the weakly interacting regime. One can see that the LOAF theory predicts a $T_{BKT}$ that agrees well with the functional form suggested by Monte Carlo simulations when $|\ln(\rho a^2)|^{-1}<0.01$.  In contrast, $T_{c}^{RG}$ seems to agree with $T_p$ when the interaction strength is extremely small. Moreover, $T_p$ and $T_{BKT}$ from the LOAF theory approach each other as $|\ln(\rho a^2)|^{-1}$ approaches zero. As the interaction becomes stronger, $T_{BKT}$ as obtained from the LOAF theory starts to deviate from $T_{BKT}$ from Eq.~\eqref{eq:TBKT_MC} and remains smaller than both $T_{c}^{RG}$ and Eq.~\eqref{eq:TBKT_MC}. The double logarithmic behavior of both $T_p$ and $T_{BKT}$ from the LOAF theory in the weakly interacting regime also suggests that our theory captures the qualitative features of both Monte Carlo and renormalization-group results.  LOAF theory serves as an extrapolation beyond weak coupling  and provides predictions that can be compared with experiments.

Interestingly, the phase diagram of Fig.~\ref{fig:2Dphase} (a) is similar to the phase diagram of a 2D Fermi gas with attractive interactions \cite{Melo2D}. There are subtle differences \cite{noML}. For example, $T_{BKT}$ for fermions increases as the attractive interactions increase but for bosons it increases as the repulsion increases. The slow increase of $T_{BKT}$ as a function of $|\ln(\rho a^2)|^{-1}$ away from the weakly interacting regime agrees qualitatively with more recent Monte Carlo simulations \cite{2DMonteCarlo}.

\subsection{Other experimental implications}
We now address the issue whether a finite amplitude of the pairing density field results in any observable effects. One observable signature is a gapped excitation energy spectrum $\omega_k=\sqrt{(\epsilon_k+\chi)^2-|A|^2}$ at finite $T$ (consistent with the analysis of Ref.~\cite{Witten2D}), which is different from the gapless Bogoliubov spectrum. The inset of Fig.~\ref{fig:2Dphase} (a) shows the gap $\omega_k(k=0)$, which is finite and increasing as $T$ increases. The dispersion may be measured using Bragg scattering \cite{Bragg_dispersion}. Another possible signature may be revealed by the analogue of radio-frequency (RF) spectroscopy, which shows the potential of measuring the spectral function, or equivalently, the imaginary part of the single-particle Green's function \cite{OurRF}. For fermions with attractive interactions, the spectral function of a homogeneous gas could show a two-peak structure due to the particle-hole mixing in the formation of Cooper pairs \cite{OurRF}. Here we investigate if there is a similar structure for bosons.

The single-particle Green's function from the LOAF theory is \cite{LOAF_SF}
\begin{equation}\label{eq:G11}
G_{MF}(k,i\omega_n)=\frac{i\omega_n+\epsilon_k+\chi}{\omega_n^{2}+\omega_{k}^{2}}.
\end{equation}
Here $\omega_n$ is the bosonic Matsubara frequency. Making the analytic continuation $i\omega_n\rightarrow\omega+i0^{+}$, one obtains $G_{MF}(k,\omega)$ \cite{Fetterbook}.  The spectral function is defined as $\mathcal{A}(k,\omega)=2\mbox{Im}G(k,\omega)$, where $G(k,\omega)$ is the single-particle Green's function of a given theory, and it satisfies the sum rule 
\begin{equation}\label{eq:sumrule}
\int_{-\infty}^{\infty}\frac{d\omega}{2\pi}\mathcal{A}(k,\omega)=1
\end{equation}
 for any $T$.

The LOAF theory predicts a qualitative difference in the spectral function when the amplitude of $A$ becomes finite. Below $T_p$, one has $\omega_k=\sqrt{(\epsilon_k+\chi+A)(\epsilon_k+\chi-A)}$ and 
$\mathcal{A}(k,\omega)=2\pi u_{k}^{2}\delta(\omega-\omega_k)-2\pi v_{k}^{2}\delta(\omega+\omega_k)$. 
Here $u_{k}^{2},v_{k}^{2}=[(\epsilon_k+\chi)/\omega_k\pm 1]/2$. This expression implies that there are two peaks at $\omega=\pm\omega_k$ for a fixed $k$. This is in contrast to the spectral function in regime I where $A=0$. In regime I, $\omega_k=\epsilon_k+\chi$ so the spectral function is $\mathcal{A}(k,\omega)=2\pi\delta(\omega-\omega_k)$. There is only one peak in the spectral function in the normal phase.

The different numbers of peaks in the spectral function below and above $T_p$ is a prediction from the LOAF theory. The delta-function peaks are due to the fact that only the leading-order contributions from the composite fields are included at the mean-field level. To obtain the widths of the spectral peaks, one has to go beyond the leading-order theory. For the relativistic $\phi^{4}$ theory it has been shown in Ref.~\cite{CooperPRD04} that by considering a self-consistent $\epsilon$ expansion for the generating functional of the two-particle irreducible graphs, one obtains an approximation to the coupled Green fuinction equations which generate a finite imaginary part in the self energy and as a consequence the spectral peaks will be broadened. 

In particular, above $T_p$ when $A=0$, our theory beyond the leading order can be formulated as a set of Schwinger-Dyson (SD) equations following Ref.~\cite{CooperPRD04} 
\begin{eqnarray}
G^{-1}&=&G^{-1}_{MF}+\Sigma, \nonumber \\
\Sigma&=&\int G \mathcal{D} \Lambda, \nonumber \\
\mathcal{D}^{-1}&=&\mathcal{D}^{-1}_{0}+\Pi, \nonumber \\
\Pi&=&\int GG\Lambda.
\end{eqnarray}
Here $\Sigma$ is the self energy of bosons, $\mathcal{D}$ is the composite-field propagator, $\Lambda$ is the interaction vertex, $D_{0}^{-1}(x,x^{\prime})=\delta(x-x^{\prime})/\lambda$ is the inverse bare composite-field propagator, $\Pi$ is the self energy of composite fields. To lowest order one may use the approximation $\Lambda=1$, $\mathcal{D}=\mathcal{D}_0$, $G=G_{MF}$, and $\Pi=\Pi_0=\int G_{MF}G_{MF}$ on the right-hand side. By writing $\Sigma=\mbox{Re}\Sigma+i\mbox{Im}\Sigma$, one can see that the delta function in the spectral function is replaced by a Lorentzian function with a full width at half maximum equal to $2\mbox{Im}\Sigma$. Below $T_p$ there are two finite composite fields, $A$ and $\chi$, and the Green's functions of $\phi$, $\chi$, and $A$ are mixed to form a $5\times 5$ matrix $G_{\alpha\beta}$, where $\alpha, \beta=1,\cdots,5$ corresponding to $(\phi,\phi^*,\chi,A,A^*)$.  Following Ref.~\cite{CooperPRD04}, the SD equations are
\begin{eqnarray}
G^{-1}_{\alpha\beta}&=&G^{-1}_{MF,\alpha\beta}+\Sigma_{\alpha\beta}, \nonumber \\
\Sigma_{\alpha\beta}&=&\int\lambda_{\alpha\alpha^{\prime}\beta^{\prime}}G_{\alpha^{\prime}\alpha^{\prime\prime}}G_{\beta^{\prime}\beta^{\prime\prime}}\Lambda_{\alpha^{\prime\prime}\beta^{\prime\prime}\beta}, \nonumber \\
\Lambda_{\alpha\beta\gamma}&=&\lambda_{\alpha\beta\gamma}-\frac{\delta\Sigma_{\alpha\beta}}{\delta\phi_{\gamma}}.
\end{eqnarray}
Here repeated indices are summed, $\lambda_{\alpha\beta\gamma}$ is the bare vertex function, and $\phi_{\gamma}$ is an element of $(\phi,\phi^*,\chi,A,A^*)$. One may implement further approximations such as the bare-vertex approximation and resort to numerical methods for evaluating the Green's function beyond the mean-field level as discussed in Ref.~\cite{CooperPRD04}.

\begin{figure}
  \includegraphics[width=3.4in,clip] {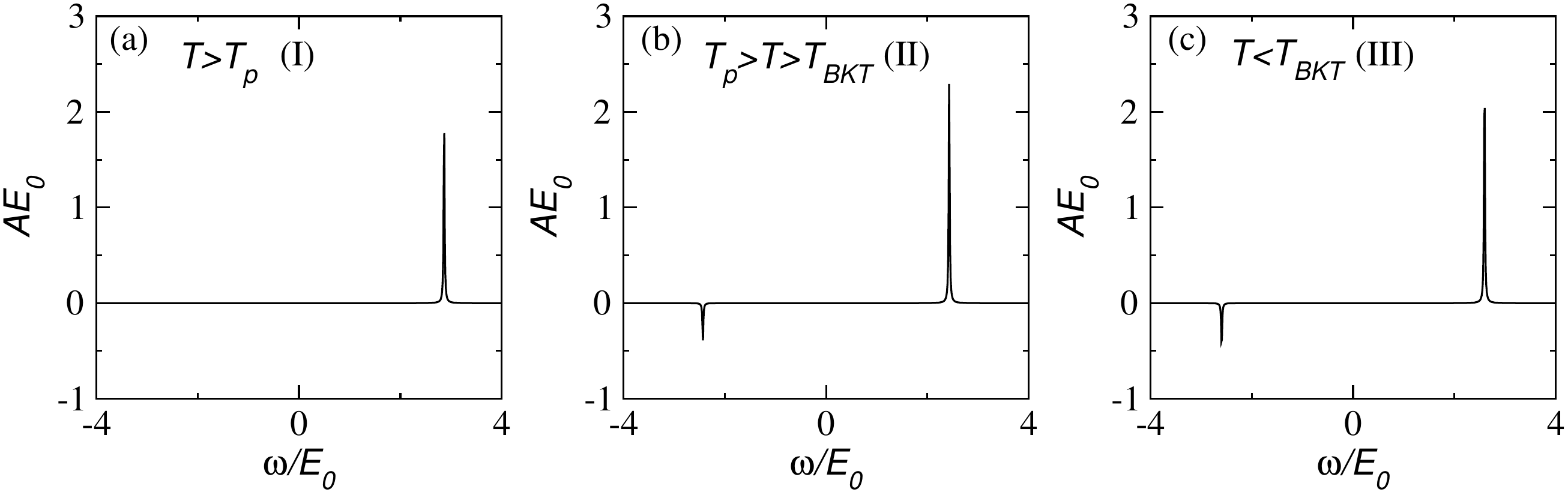}
  \caption{Spectral functions at fixed $k=k_0$ for (a) $T>T_p$ ($T/T_0=8$), (b) $T_p>T>T_{BKT}$ ($T/T_0=4$), and (c) $T<T_{BKT}$ ($T/T_0=1.3$). They belong to regimes I, II, and III of Fig.~\ref{fig:2Dphase} (a), respectively. Here $|\ln(\rho a ^2)|^{-1}=0.5$.}
\label{fig:Akw}
\end{figure} 
Instead of performing these intricate  numerical calculations that should only lead to quantitative corrections, here we focus on the qualitative features of the spectral function already present in the mean-field level but modified by the broadening of the higher-order corrections. The delta functions in $\mathcal{A}(k,\omega)$ evaluated from $G_{MF}$ will be broadened by effects beyond the LOAF theory and for illustrative purposes only, we  introduce a Lorentzian function by $\delta(x)\rightarrow(\frac{1}{\pi})\frac{\Gamma/2}{x^2+(\Gamma/2)^2}$ to approximate the spectral peaks. The width $\Gamma$ could be obtained from the full expression of the self energy $\Sigma$ outlined above, but here for simplicity we set $\Gamma/E_0=0.02$. A fully numerical calculations similar to the $T=0$ calculations in Ref.~\cite{FRG2D} may help determine the dependence of $\Gamma$ on $T$ and the interaction strength. Our approximation of the delta function still respects the sum rule \eqref{eq:sumrule} for $\mathcal{A}(k,\omega)$. Figure~\ref{fig:Akw} shows the spectral function $\mathcal{A}(k,\omega)$ at fixed $k=k_0$ for three selected temperatures corresponding to regimes I, II, and III of Fig.~\ref{fig:2Dphase}. 

When $T>T_p$, $A=0$ so there is only one peak as shown in Fig.~\ref{fig:Akw}(a). Below $T_p$ the finite amplitude of the pairing density field induces another peak in the negative energy region. In the Bogoliubov theory of weakly interacting bosons, the Bogoliubov transformation mixes the creation and annihilation operators. Since the LOAF theory is a natural generalization of the Bogoliubov theory, the pairing density field includes similar mixing effects. Thus the spectral weights of positive and negative energy states are correlated. The two-peak structure  reflects this type of correlation effects. The spectral function of a $T=0$ 2D Bose gas has been evaluated using numerical functional renormalization group method focusing on the positive-energy peak \cite{FRG2D}. Since the single-particle Green's function \eqref{eq:G11} only contains information about the amplitude of the pairing density field, it does not exhibit observable signatures of the BKT transition (see Fig.~\ref{fig:Akw} (b) and (c)). 

We note that the spectral function of a fermionic BCS superfluid is $\mathcal{A}(k,\omega)=\tilde{u}_{k}^{2}\delta(\omega-E_k)+\tilde{v}_{k}^{2}\delta(\omega+E_k)$, where $\tilde{u}_{k}^{2},\tilde{v}_{k}^{2}=[1\pm(\epsilon_k-\mu)/E_k]/2$, $E_{k}=\sqrt{(\epsilon_k-\mu)^2+\Delta^2}$, and $\Delta$ is the gap function. When $\Delta >0$, there are two \textit{positive} peaks reflecting the pairing between fermions. The spin statistics, nevertheless, causes one positive and one \textit{negative} peaks for bosons. 
%One can also see this difference from the sum rule of $\mathcal{A}(k,\omega)$: $\tilde{u}_{k}^{2},\tilde{v}_{k}^{2}\le 1$ so there can be two positive peaks while $u_{k}^{2}>1$ so there must be a negative peak with the weight $v_{k}^{2}$ to satisfy the sum rule \eqref{eq:sumrule}. 
We emphasize that although this negative peak should also survive in 3D Bose gases \cite{Bogo_note}, its appearance in 2D Bose gases is a more direct evidence of the many-body pairing effect because the BEC vanishes at finite $T$.

Spectroscopies probing single-particle excitations such as RF measurements may be sensitive only to the existence of an energy gap but not to phase coherence \cite{OurRF}. From Fig.~\ref{fig:2Dphase} (b) and Fig.~\ref{fig:Akw} we reach a similar conclusion. To probe the BKT transition and the superfluid phase below it, in addition to Ref.~\cite{HadzibabicBKT} we suggest experiments such as the measurement of the second sound, which has been shown to be an indication of superfluidity using hydrodynamic approaches in both 3D \cite{Oursound} and quasi-1D \cite{Stringarisound} geometries and should have the same resolution in 2D.

\section{Conclusion}\label{sec:conclusion}
In summary, we present a coherent mean-field picture of pairing effects, superfluidity, BKT physics, and single-particle excitations by introducing phase fluctuations into the LOAF theory of a 2D interacting Bose gas in a manner similar to what is done in the 2D BCS theory as well as in the Thirring model. In addition to mapping out the phase diagram at finite $T$, our theory predicts observable signatures of pairing effects above the BKT transition temperature, which resembles the pseudogap physics of 2D Fermi gases \cite{KohlPG}. The LOAF theory agrees reasonably with previous results from perturbative calculations, renormalization-group calculations, and Monte Carlo simulations in the weakly interacting regime but further explores the regime of intermediate interaction strength.  By implementing the  local density approximation for trapped gases, our theory may provide more insights into experiments such as Refs.~\cite{Hung2D,Yefsah2D}.

The authors acknowledge the support of the U. S. DOE through the LANL/LDRD Program. We thank Santa Fe Institute for its hospitality.

\bibliographystyle{apsrev4-1}
%\bibliography{reference} 

\begin{thebibliography}{46}
\bibitem [{noM()}]{noML}%
  \BibitemOpen
  \href@noop {} {}\bibinfo {note} {By considering higher-order terms of the
  action depending on $\theta$, it is possible to include the
  melted-vortex-lattice phase discussed in Ref.~\cite{Melo2D}.}\BibitemShut
  {Stop}%
\bibitem [{Bog()}]{Bogo_note}%
  \BibitemOpen
  \href@noop {} {}\bibinfo {note} {In the presence of BEC, there should be one
  more peak at $\omega=0$ due to BEC.}\BibitemShut {Stop}%
\end{thebibliography}
%merlin.mbs 2010-03-15 4.21a (PWD, AO, DPC)

\end{document}